\documentclass[12pt]{article}
\usepackage{graphicx}
\begin{document}
\begin{titlepage}

\title{Plane Gravitational Waves, the Kinetic Energy of Free Particles
and the Memory Effect}

\author{J. W. Maluf$\,^{(1)}$, J. F. da Rocha-Neto$\,^{(2)}$, \\
S. C. Ulhoa$\,^{(3)}$, and F. L. Carneiro$\,^{(4)}$ \\
Instituto de F\'{\i}sica, \\
Universidade de Bras\'{\i}lia\\
70.919-970 Bras\'{\i}lia DF, Brazil\\}
\date{}
\maketitle
\bigskip
\bigskip

\begin{abstract}
It is shown that in the passage of a short burst of non-linear plane 
gravitational wave, the kinetic energy of free particles may either 
decrease or increase. The decreasing or increasing of the kinetic energy 
depends crucially on the initial conditions (position and velocity) of the
free particle. Therefore a plane gravitational wave may extract energy from
a physical system.
\end{abstract}
\thispagestyle{empty}
\vfill
\par
\bigskip

\bigskip
{\footnotesize
\noindent (1) wadih@unb.br, jwmaluf@gmail.com\par
\noindent (2) rocha@fis.unb.br\par
\noindent (3) sc.ulhoa@gmail.com\par
\noindent (4) fernandolessa45@gmail.com\par}

\end{titlepage}
\newpage

\section{Introduction}
The memory effect is a very interesting feature of gravitational waves. 
It expresses the change in a physical system (a detector), between the
final and initial state of the system, after the passage of a gravitational
wave. It is an effect that might be observable in the future LISA operations,
and also raises quite interesting theoretical issues. If the physical system
is composed of free test particles, the memory effect is determined by the 
permanent displacement of the particles, caused by the gravitational wave.
The memory effect due to bursts of plane gravitational waves has been recently
considered by Zhang, Duval, Gibbons and Horvathy \cite{ZDGH1,ZDGH2}. 
Zel´dovich and Polnarev  \cite{ZP} first considered the effect of linearised
gravitational waves on non-interacting bodies, such as satellites, but the 
memory effect seems to be first proposed by Braginsky and Grishchuk \cite{BG},
who were really interested in the motion of free particles in the space-time 
of a gravitational wave. 
Soon after, a distinction was made between gravitational wave bursts with 
and without memory \cite{BT}. This distinction was already considered in Ref.
\cite{GH}, but not in the context of the memory effect. A non-linear form of 
memory effect had been discovered independently by Blanchet and Damour 
\cite{BD}, and by Christodoulou \cite{Christodoulou}. A thorough mathematical
treatment of the memory effect has been recently made by Favata \cite{Favata}.

One interesting manifestation of the memory effect is the velocity memory 
effect, that is characterized by a permanent change of the velocity of the free
particle after the passage of the wave. This effect was considered by Souriau 
\cite{Souriau}, Braginsky and Thorne, \cite{BT}, Grishchuk and Polnarev
\cite{GP}, Bondi and Pirani \cite{BP}, and more recently by Zhang, Duval and
Horvathy \cite{ZDH}. 

In this article, we consider the approach to plane gravitational waves given in
Ref. \cite{ZDGH2}, and address the memory effect from the point of view of the
velocities of free particles, after the passage of a wave.
The wave burst is modelled by Gaussian functions of the 
retarded time. If the free particles are initially at rest in the space-time, 
then they acquire velocity after the passage of the wave. Physically, we expect
a tiny variation of the velocity of the particles, because the gravitational
wave is supposed to be very weak. However, we have found that for certain 
initial conditions of the free particles, with non-vanishing initial velocities,
the final kinetic energy of the particles is smaller than the initial energy. 
Therefore, contrary to the common expectation, the gravitational wave may absorb
energy of the physical system. We argue that such extraction of energy is 
consistent with the expression of the gravitational energy-momentum of 
plane-fronted gravitational waves calculated in Ref. \cite{Maluf1}, and might
explain the propagation of waves with very slow dissipation in time.

\section{Free particles and gravitational waves}

A non-linear plane gravitational wave may be written in several different forms
\cite{Kramer}. One possible form is the plane-fronted gravitational wave that 
travels in the $z$ direction, and which is given by \cite{Ehlers,Ehlers-2,HS}

\begin{equation}
ds^2=dx^2+dy^2+2du\,dv+H(x,y,u)du^2\,.
\label{1}
\end{equation}
The function $H(x,y,u)$ satisfies

\begin{equation}
\biggl( {\partial^2 \over {\partial x^2}}+
{\partial^2 \over {\partial y^2}}\biggr)H(x,y,u)=0\,.
\label{2}
\end{equation}
Transforming $(u,v)$ to $(t,z)$ coordinates, where

\begin{equation}
u={1\over \sqrt{2}}(z-t)\,, \ \ \ \  v={1\over \sqrt{2}}(z+t)\,,
\label{2-1}
\end{equation}
we find

\begin{equation}
ds^2=\biggl({H\over 2} -1\biggr)dt^2+dx^2+dy^2+
\biggl({H\over 2}+1\biggr) dz^2-H\,dt dz\,.
\label{3}
\end{equation}
We are assuming $c=1$.
The function $H$ must satisfy only Eq. (2). Otherwise it is arbitrary,
specially regarding the dependence on the retarded time $(-u)$. 
The geodesic equations in terms of the $t, x, y, z$ coordinates are,
respectively \cite{JF}

\begin{equation}
2\ddot{t} + \sqrt{2}H\ddot{u} + \sqrt{2}\dot{H}\dot{u} 
- {1\over \sqrt{2}}{\partial H\over \partial u}\dot{u}^2 = 0,
\label{3-4a}
\end{equation}
\begin{equation}
2\ddot{x} - {\partial H\over \partial x}\dot{u}^{2} = 0,
\label{3-4b}
\end{equation}
\begin{equation}
2\ddot{y} - {\partial H\over \partial y}\dot{u}^{2} =  0,
\label{3-4c}
\end{equation}
\begin{equation}
2\ddot{z} +  \sqrt{2}H\ddot{u} + \sqrt{2}\dot{H}\dot{u} 
- {1\over \sqrt{2}}{\partial H\over \partial u}\dot{u}^2 = 0\,,
\label{3-4d}
\end{equation}
where the dot represents derivative with respect to $s$.
From the first and fourth equations above we get

\begin{equation}
\ddot{z} - \ddot{t} = 0 \to \ddot{u} = 0 \to \dot{u}=
{1\over \sqrt{2}}(\dot{t} - \dot{z}) = constant.
\label{3-4e}
\end{equation}

The line element for the gravitational wave considered in Refs. 
\cite{ZDGH1,ZDGH2} is presented in Brinkmann coordinates \cite{Brinkmann},
and is similar to Eq. (\ref{1}). In terms of the notation above, the 
line element for a wave propagating in the $z$ direction is given by

\begin{equation}
ds^2=dx^2+dy^2+2du\,dv+K_{ij}(u)x^ix^j du^2\,,
\label{4}
\end{equation}
where

\begin{equation}
K_{ij}(u)\,x^i x^j={1\over 2}A_{+}(u)(x^2-y^2)+A_{\times}(u)\,xy\,.
\label{5}
\end{equation}
In the expression above, $(+\,,\times)$ represent the two polarization states.
There are several possible choices for the amplitude $A_{(+,\times)}$, as 
discussed in Ref. \cite{ZDGH2}. These choices are given by a Gaussian and by
derivatives of the Gaussian. We will choose to work with a simple Gaussian,

\begin{equation}
A_{(+,\times)}(u)={1\over L^2}\, e^{-u^2 /\lambda^2}\,,
\label{6}
\end{equation}
where $L$ and $\lambda$ are constants with dimension of length: $\lambda$ is
related to the width of the Gaussian, and $L^2$ could be interpreted as the 
size of the transversal area of the wave (we are requiring $g_{00}<0$ in 
Eq. (\ref{3})). These constants are necessary, so that
the line element (4) has dimension of (length)$^2$.
The results to be presented below do not depend on whether we use a Gaussian 
or derivatives of the Gaussian, as in Ref. \cite{ZDGH2}

In what follows, we 
will consider three simple choices for $A_{+}$: $A_{+}=e^{-u^2}$, 
$A_{+}=(1/4)e^{-u^2/2}$ and $A_{+}=(1/8)e^{-u^2/3}$. We must remember that the
constants $L$ and $\lambda$ are present in the expression of $A_{+}$ and, for
our purposes, it makes no difference whether we use centimetres or metres.

Free particles follow geodesics in space-time. The geodesics in the space-time 
of a wave determined by

\begin{equation}
K_{ij}(u)\,x^i x^j={1\over 2}A_{+}(u)(x^2-y^2)\,,
\label{7}
\end{equation}
satisfy the equations 

\begin{eqnarray}
{{d^2 x}\over du^2}&-&{1\over 2}A_{+}\,x=0\,, \\ 
{{d^2 y}\over du^2}&+&{1\over 2}A_{+}\,y=0\,, \\
{{d^2 v}\over du^2}&+& {1\over 4}{{dA_{+}}\over du} (x^2-y^2)+
A_{+}\biggl(x{{dx}\over du}-y{{dy}\over du}\biggr)=0\,,
\label{8}
\end{eqnarray}
which are strictly equivalent to eqs. (\ref{3-4a}-\ref{3-4d}) provided we make 
$s=u$ and identify

$$H={1\over 2}A_{+}(u)(x^2-y^2)\,.$$

By using the program MAPLE 12 we have specified initial conditions for a free
particle and solved the equations above numerically for the velocities
$V_x=dx/du$, $V_y=dy/du$ and $V_z=dz/du$. Then we have evaluated 

$$2K={1\over 2}(V_x^2 + V_y^2+ V_z^2)\;,$$
before and after the passage of the wave. $K$ is the kinetic energy per unit 
mass, given by 

$$K={1\over 2}\biggl[ \biggl({{dx}\over {dt}}\biggr)^2+ 
\biggl({{dy}\over {dt}}\biggr)^2+\biggl({{dz}\over {dt}}\biggr)^2 \biggr]\,.$$
The results are displayed in the figures below. 

The initial conditions for all geodesic curves, at $u=0$, are
$x(0)=1$, $y(0)=1$, $z(0)=0$ and $V_z(0)=0$. The {\bf increasing} of the kinetic
energy per unit mass occurs with the initial conditions

\begin{equation}
V_x(0)=0.4\;,  \ \ \ \ V_y(0)=0\,,
\label{9}
\end{equation}
and the {\bf decreasing} of the kinetic energy takes place for the initial 
conditions

\begin{equation}
V_x(0)=0\;, \ \ \ \ V_y(0)=0.4\,.
\label{10}
\end{equation}

The three cases we will consider are\par
\bigskip

\noindent Case 1: $A_{+}=e^{-u^2}$\par
\bigskip
\noindent Case 2: $A_{+}={1\over 4}e^{-u^2/2}$\par
\bigskip
\noindent Case 3: $A_{+}={1\over 8}e^{-u^2/3}$\par 
\bigskip
It is important to note that, in view of the definition
$u={1\over \sqrt{2}}(z-t)$, the time coordinate $t$ runs in the opposite 
direction relatively to $u$. Therefore, since the coordinate $u$ runs from 
left to right in the figures, {\bf $t$ runs from right to left}.

\begin{figure}[h]
\centering
\includegraphics[width=0.50\textwidth]{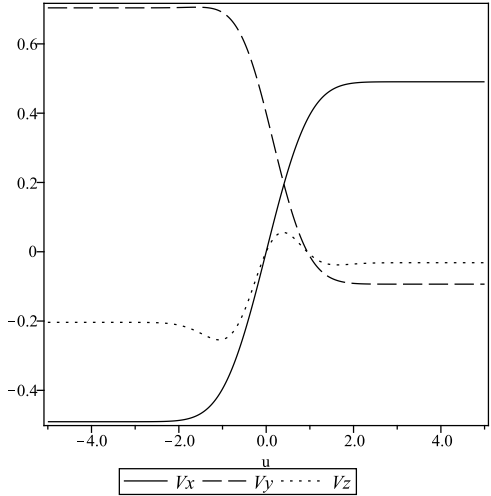}
\caption{Case 1 - velocities before and after the passage of the wave for 
the initial conditions (\ref{9})}
\end{figure}

\begin{figure}[h]
\centering
\includegraphics[width=0.50\textwidth]{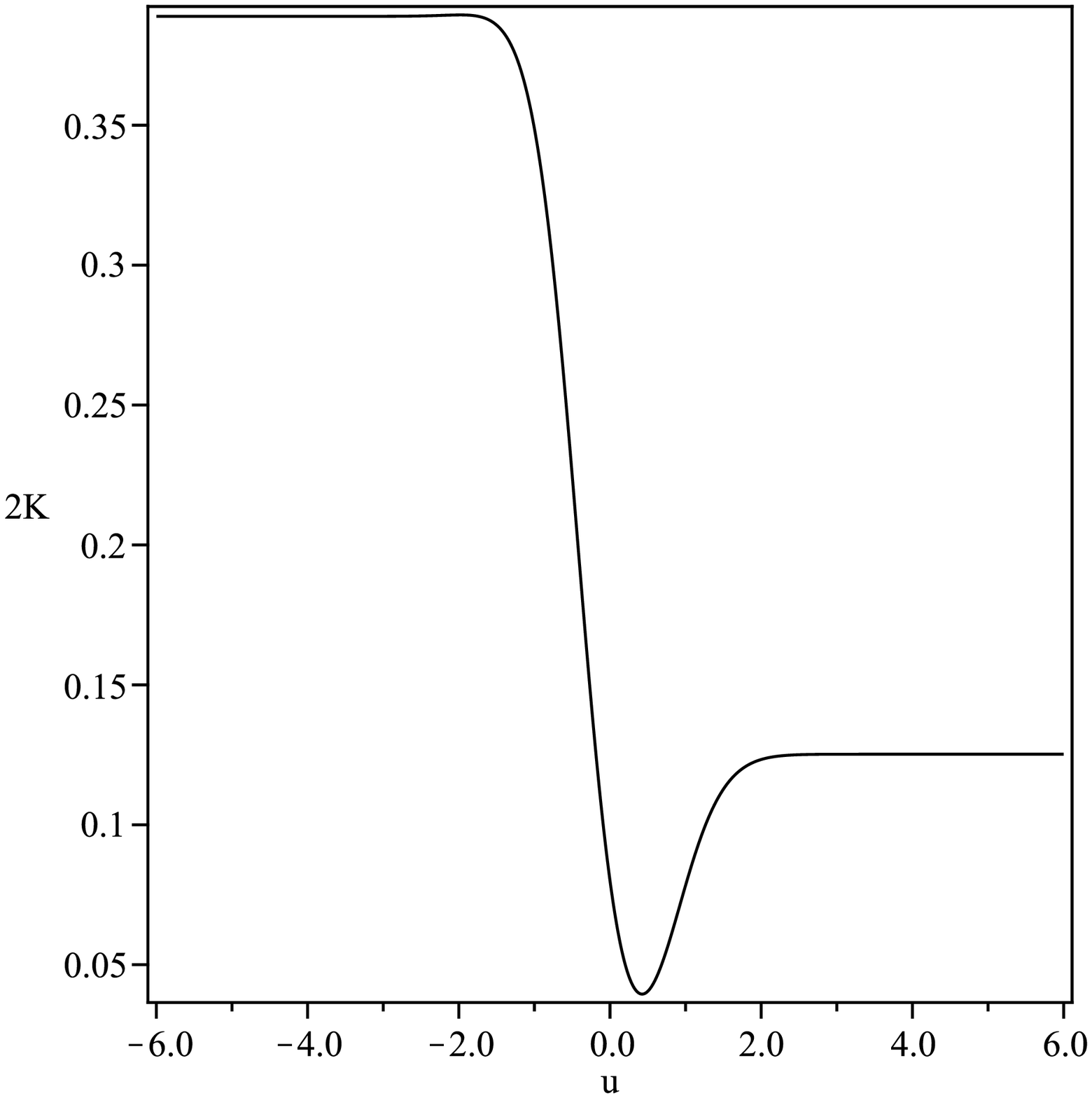}
\caption{Case 1 - increasing of the kinetic energy - initial conditions (\ref{9})}
\end{figure}

\begin{figure}[h]
\centering
\includegraphics[width=0.50\textwidth]{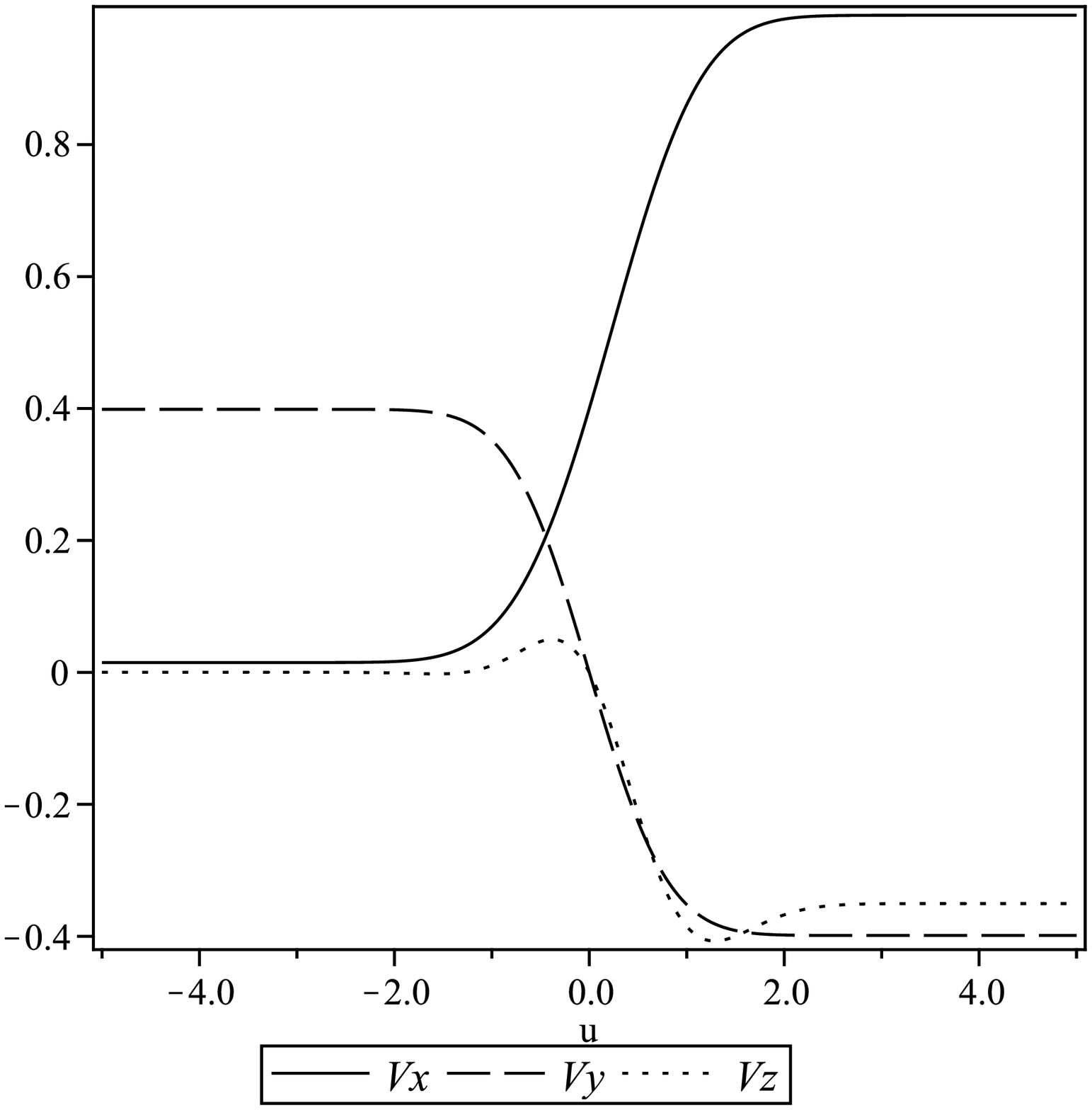}
\caption{Case 1 - velocities before and after the passage of the wave for the
initial conditions (\ref{10})}
\end{figure}

\begin{figure}[h]
\centering
\includegraphics[width=0.50\textwidth]{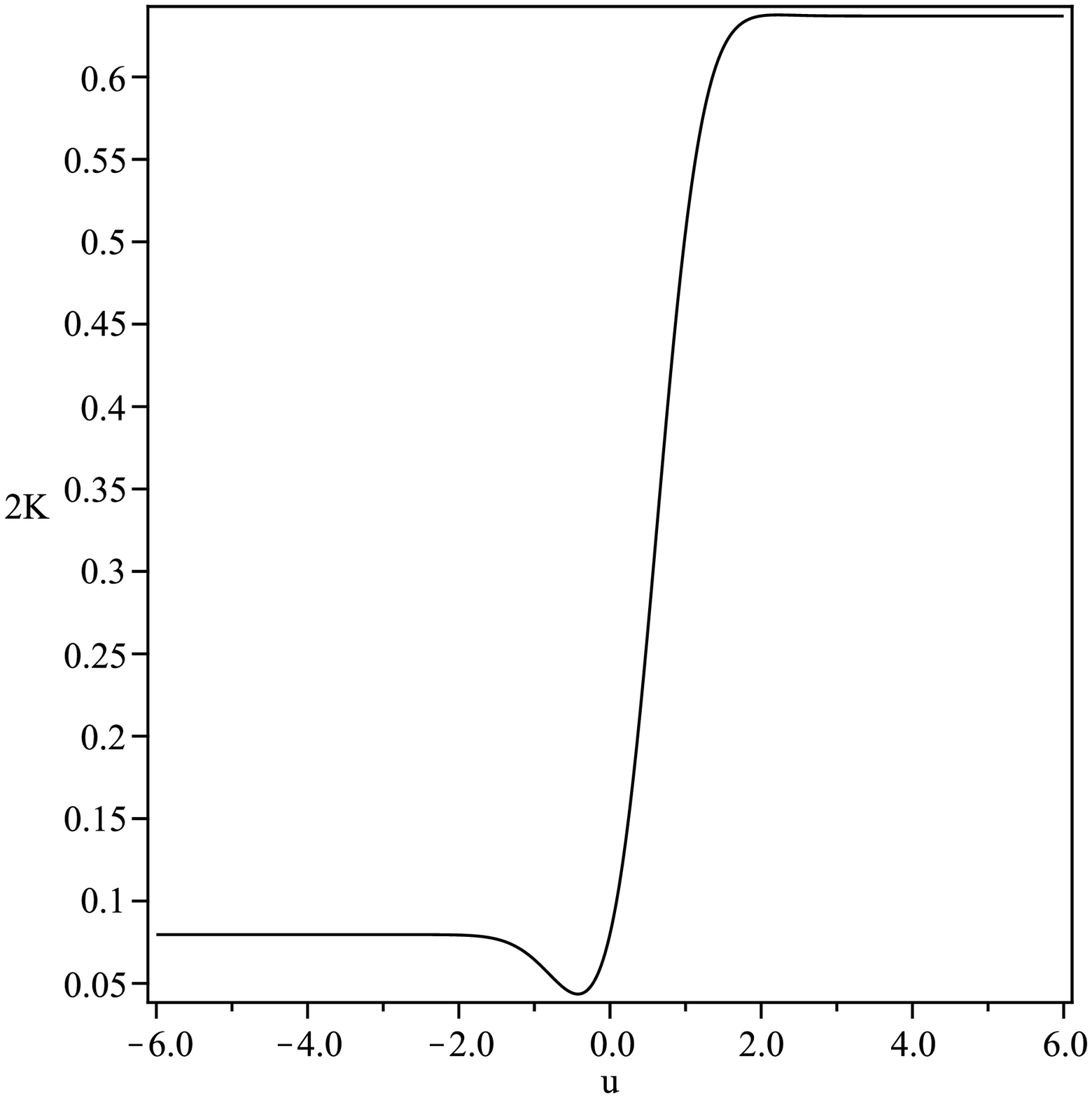}
\caption{Case 1- decreasing of the kinetic energy - initial conditions 
(\ref{10})}
\end{figure}

Figures (4), (8) and (12) clearly show the decreasing of the kinetic energy 
per unit mass of the particle, after the passage of the gravitational wave. 
The local space-time is flat before and after the passage of the wave.

Other initial conditions on the free particle lead to similar 
results, including the situation in which the kinetic energy is not changed.
We have not found a criterium for choosing general initial conditions such that
the kinetic energy always increases or always decreases. The result also does 
not depend on using the polarization $A_{\times}$. Taking into account the 
latter, the qualitative results are the same, and
the kinetic energy might likewise increase or decrease.

\section{The energy-momentum of plane-fronted gravitational waves}

The gravitational energy momentum of the plane-fronted gravitational wave 
described by Eq. (\ref{3}) has been calculated in Ref. \cite{Maluf1}, in the 
context of the teleparallel equivalent of general relativity (TEGR) 
\cite{Maluf2}. The latter is a tetrad description of the gravitational field, 
and the gravitational energy-momentum has been evaluated in the frame of a 
static observer in space-time. In the TEGR, the gravitational energy-momentum
and the 4-angular momentum satisfy the algebra of the Poincar\' e group.
We will not repeat the details of the calculations here, but just present the
final result. We have found that the non-vanishing components of the
gravitational energy-momentum $P^a$ for the wave that travels in the $z$
direction, given by Eq. (\ref{3}), are given by

\begin{equation}
P^{(0)}=P^{(3)}=-{k\over 8}\int_V d^3x
{{(\partial_i H)^2}\over {(-g_{00})^{3/2}}} \leq 0\,,
\label{13}
\end{equation}
where $k=c^3/16\pi G=1/16\pi$, $g_{00}= -1+H/2$ and $i=(x,y)$.
The expression of $P^a$ satisfies $P^aP^b\eta_{ab}=0$,
where $\eta_{ab}=(-1,+1,+1,+1)$. $P^{(0)}$ and $P^{(3)}$ are evaluated
over an arbitrary volume $V$ of the three-dimensional space. 

The gravitational energy $P^{(0)}$ is non-positive.
Therefore we see that in order for a gravitational wave given by Eqs. (\ref{1})
or (\ref{3}) to dissipate in space, it must absorb positive energy from the 
medium where it travels. Of course, the wave also transfers positive energy to 
the physical system, as we have seen, and the occurrence of both processes 
might explain why a gravitational wave travels in space for periods of time 
such as billions of years, without dissipating.

\begin{figure}[h]
\centering
\includegraphics[width=0.50\textwidth]{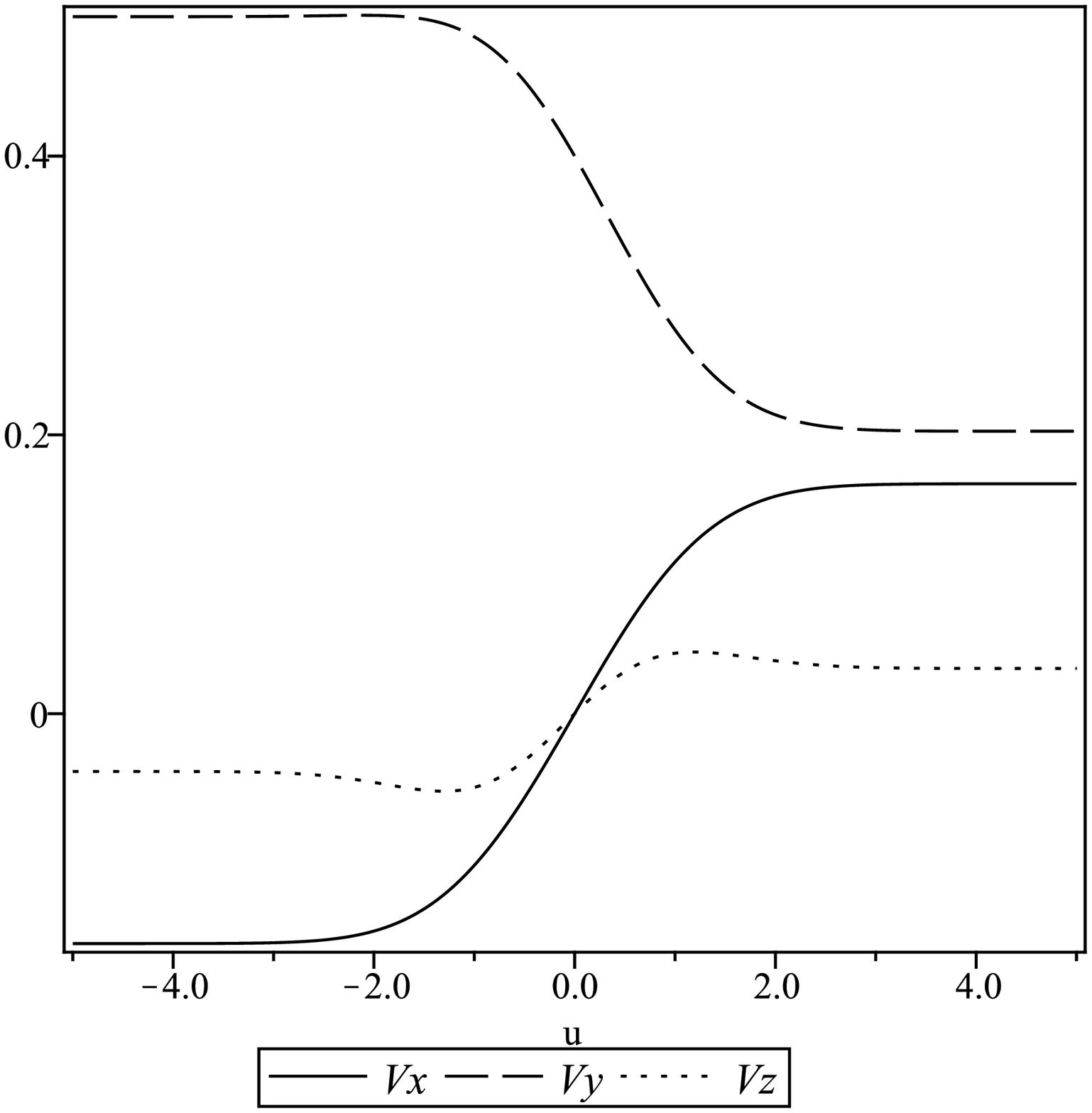}
\caption{Case 2 - velocities before and after the passage of the wave for the 
initial conditions (\ref{9})} 
\end{figure}

\begin{figure}[h]
\centering
\includegraphics[width=0.50\textwidth]{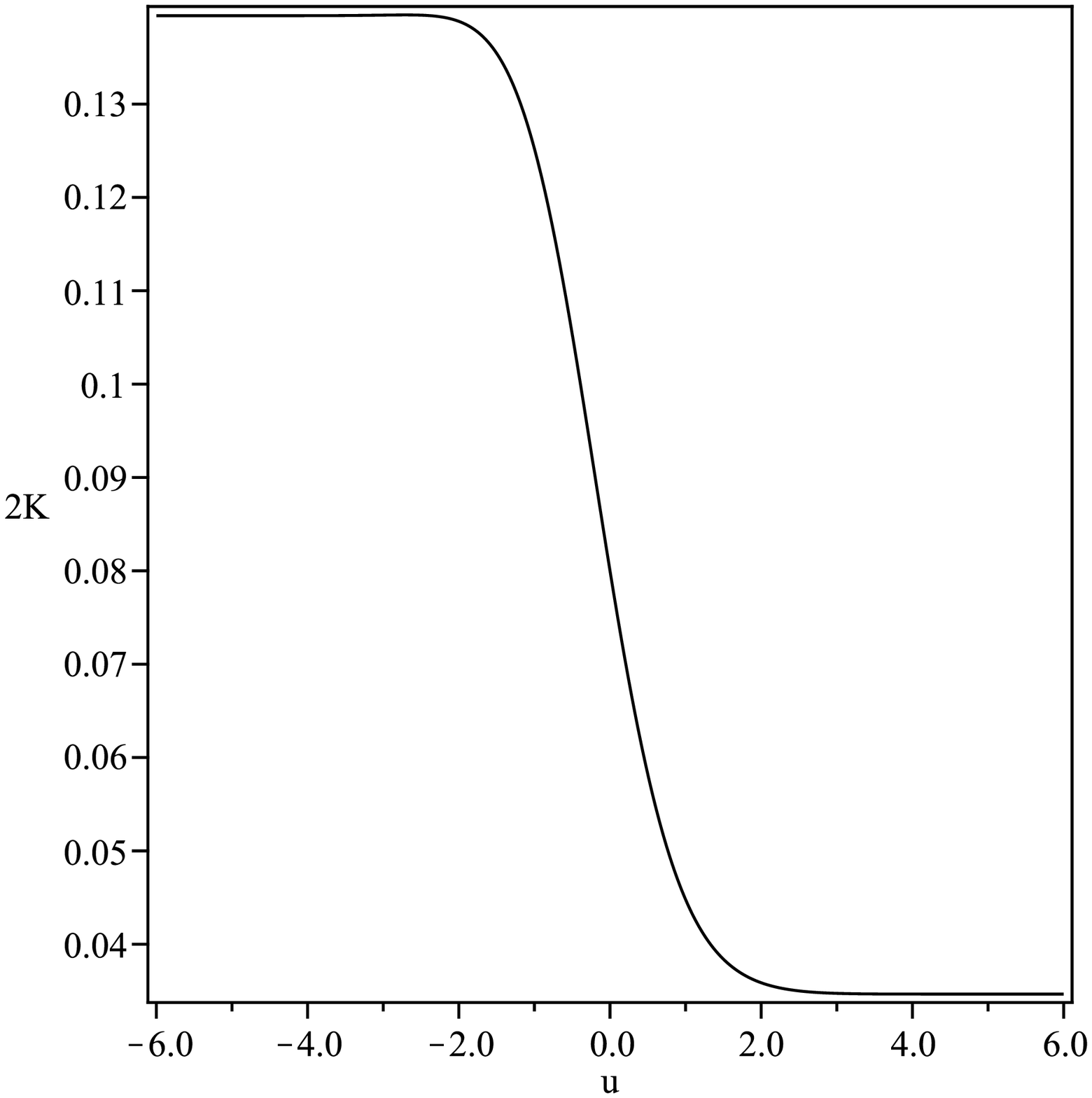}
\caption{Case 2 - increasing of the kinetic energy - initial conditions 
(\ref{9})}
\end{figure}

\begin{figure}[h]
\centering
\includegraphics[width=0.50\textwidth]{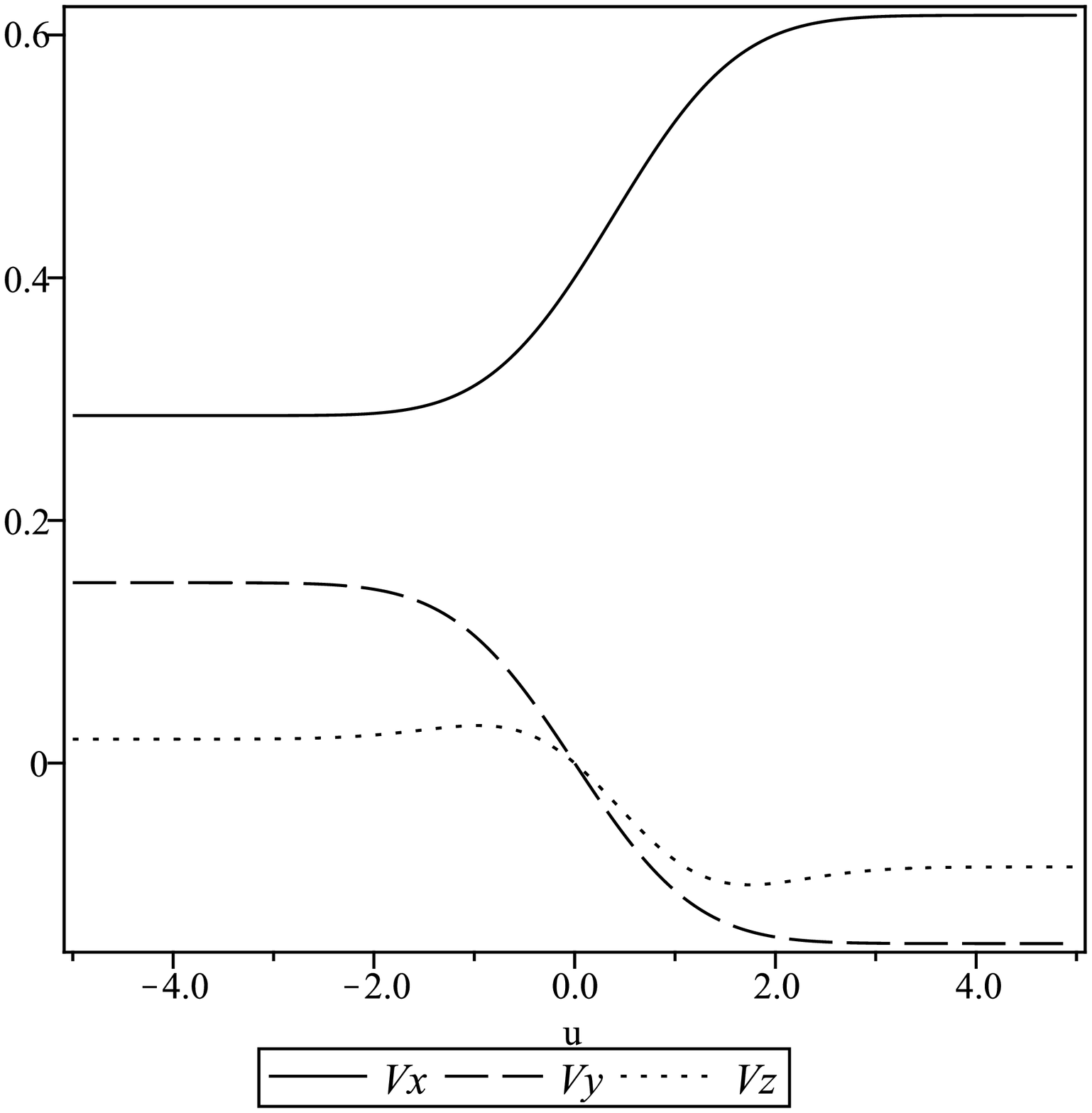}
\caption{Case 2 - velocities before and after the passage of the wave for the
initial conditions (\ref{10})}
\end{figure}

\begin{figure}[h]
\centering
\includegraphics[width=0.50\textwidth]{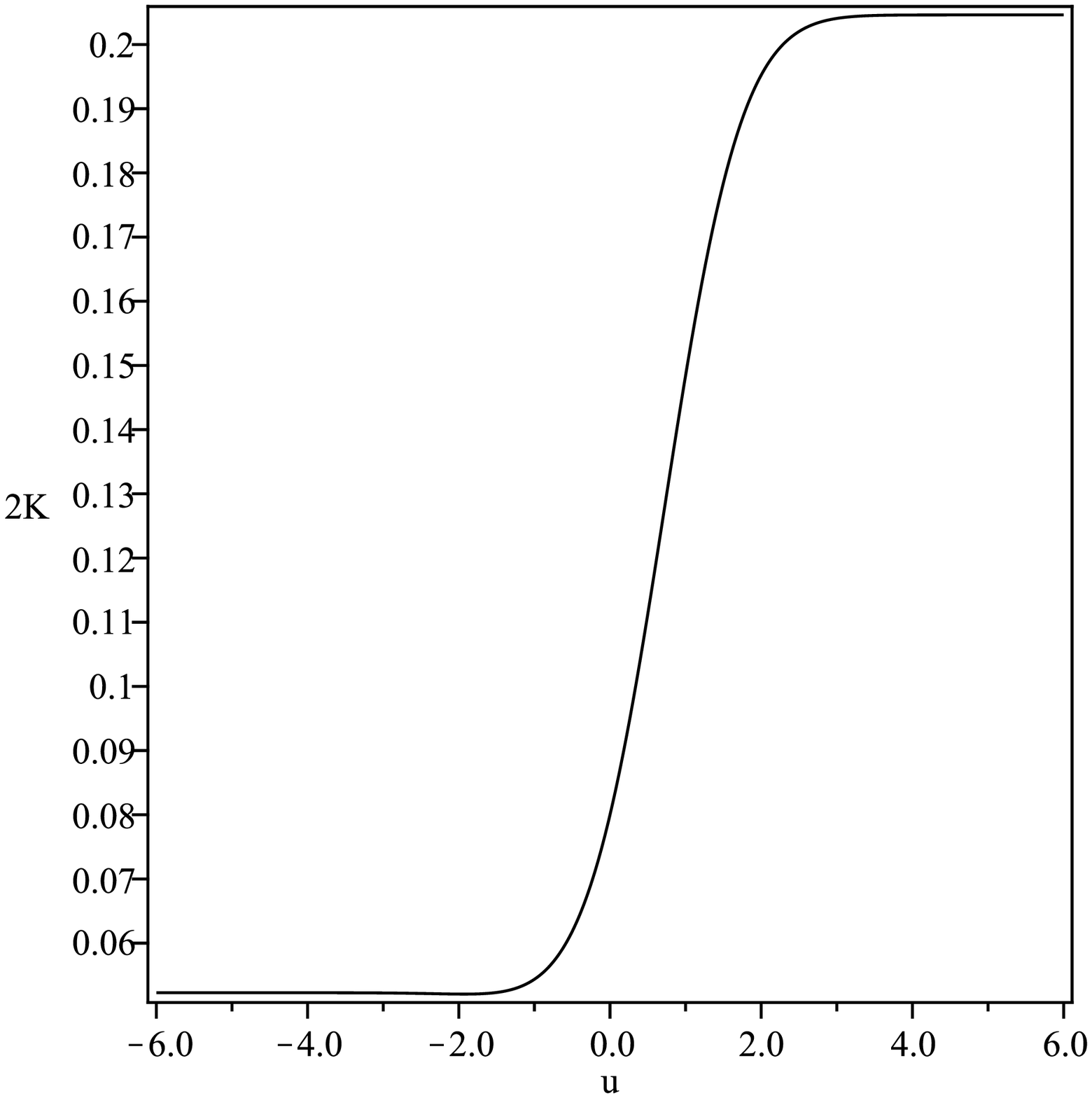}
\caption{Case 2- decreasing of the kinetic energy - initial conditions 
(\ref{10})}
\end{figure}

\begin{figure}[h]
\centering
\includegraphics[width=0.50\textwidth]{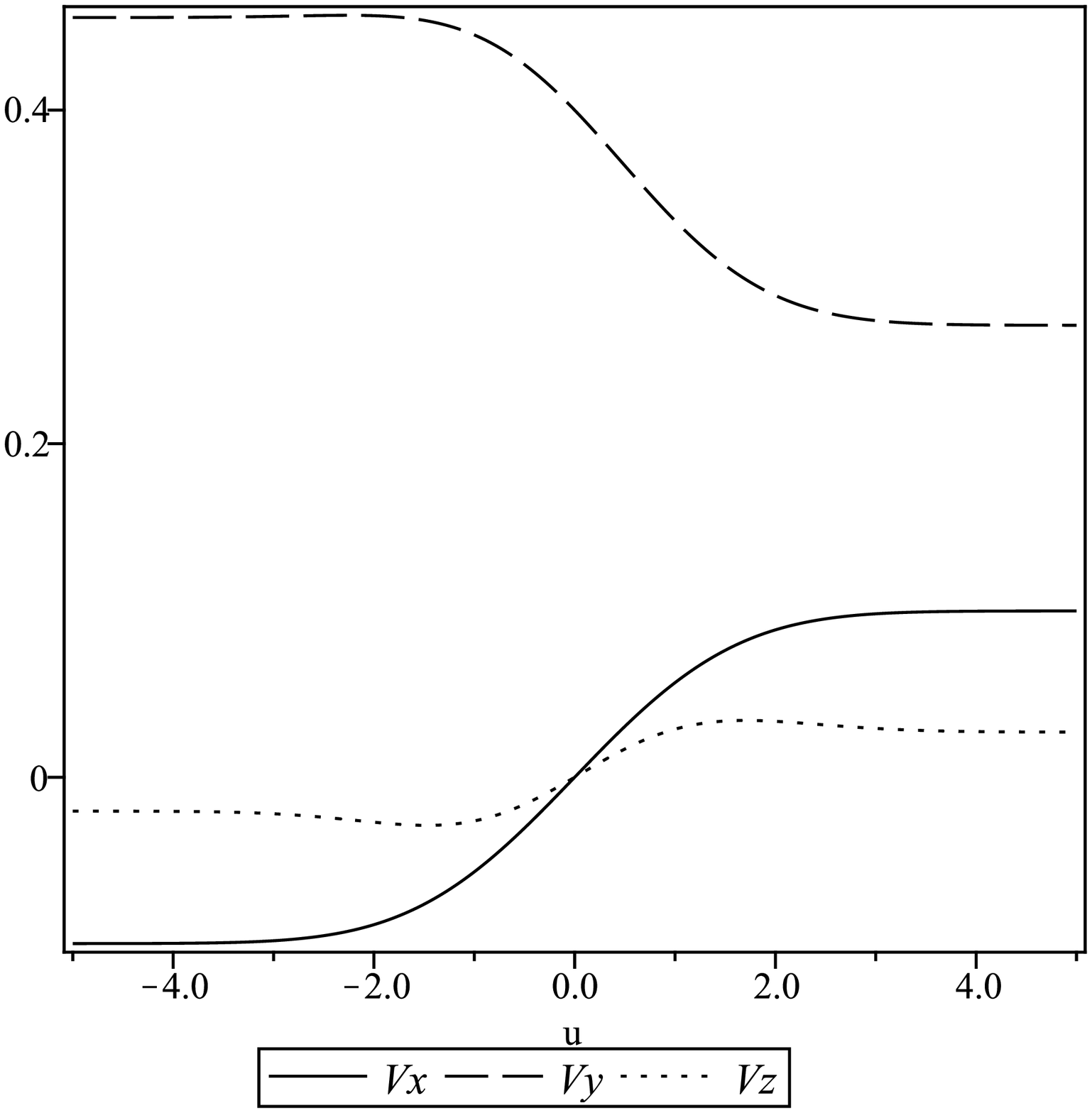}
\caption{Case 3 - velocities before and after the passage of the wave for the 
initial conditions (\ref{9})} 
\end{figure}

\begin{figure}[h]
\centering
\includegraphics[width=0.50\textwidth]{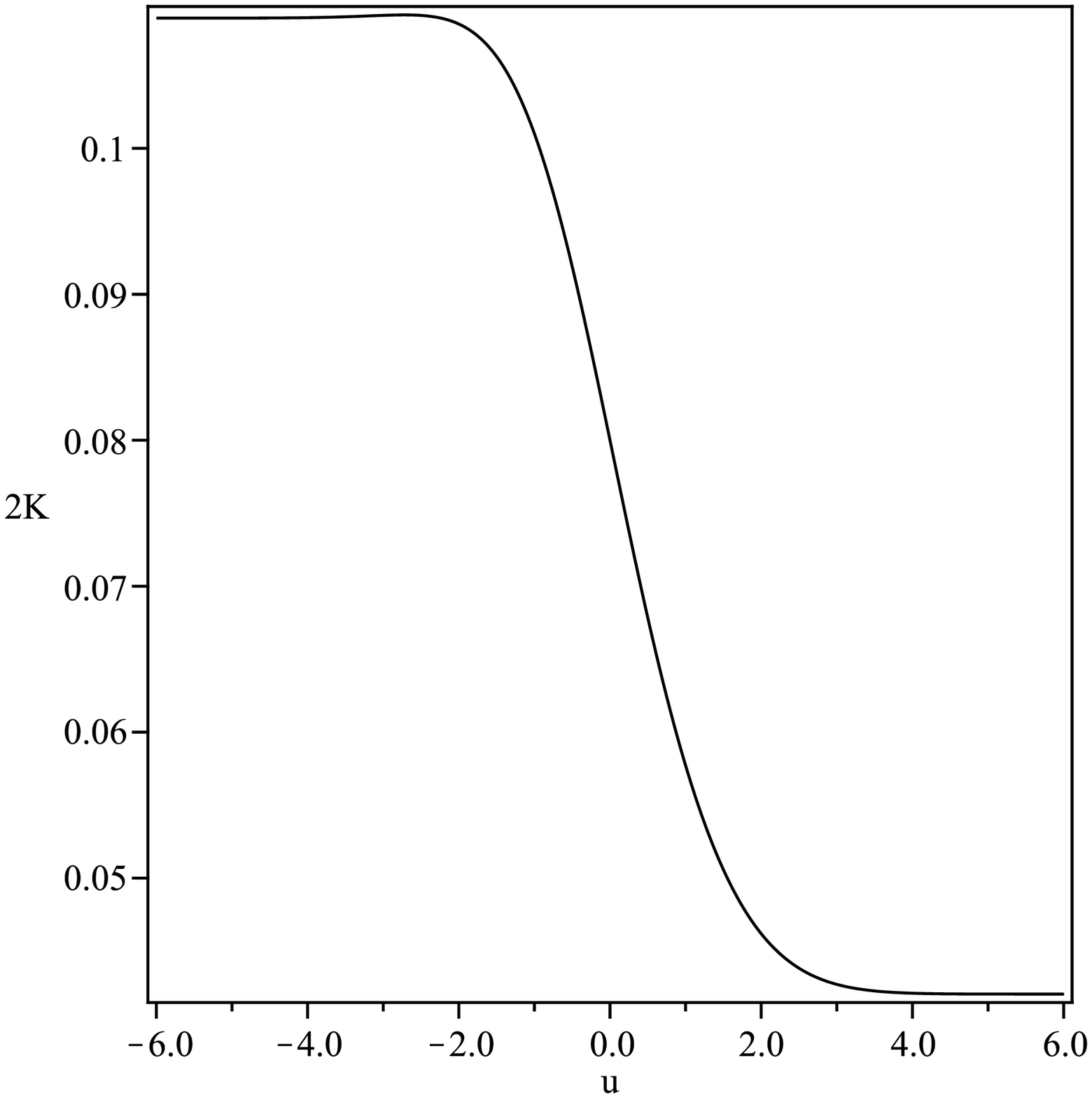}
\caption{Case 3 - increasing of the kinetic energy - initial conditions 
(\ref{9})}
\end{figure}

\begin{figure}[h]
\centering
\includegraphics[width=0.50\textwidth]{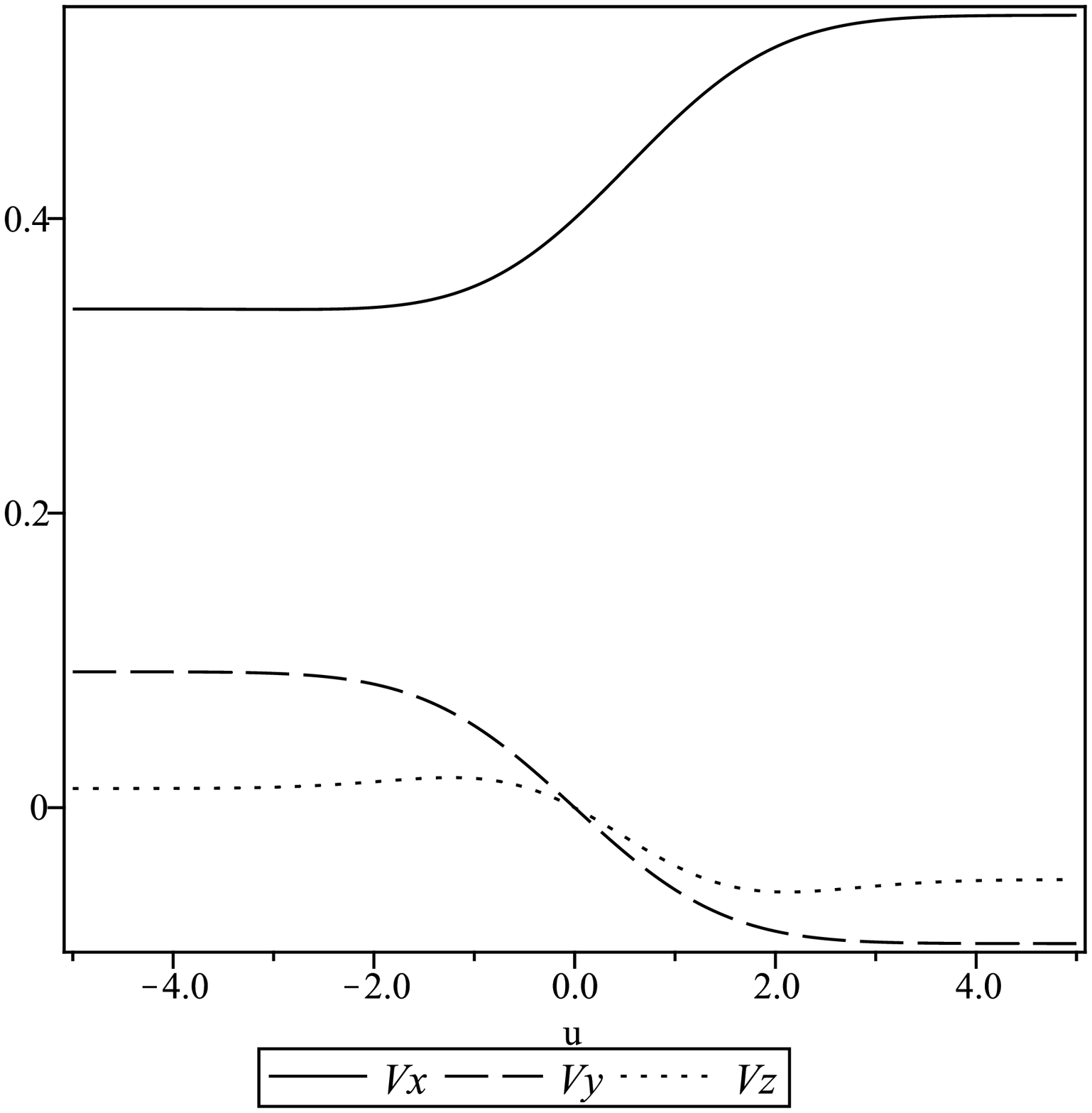}
\caption{Case 3 - velocities before and after the passage of the wave for the
initial conditions (\ref{10})}
\end{figure}

\begin{figure}[h]
\centering
\includegraphics[width=0.50\textwidth]{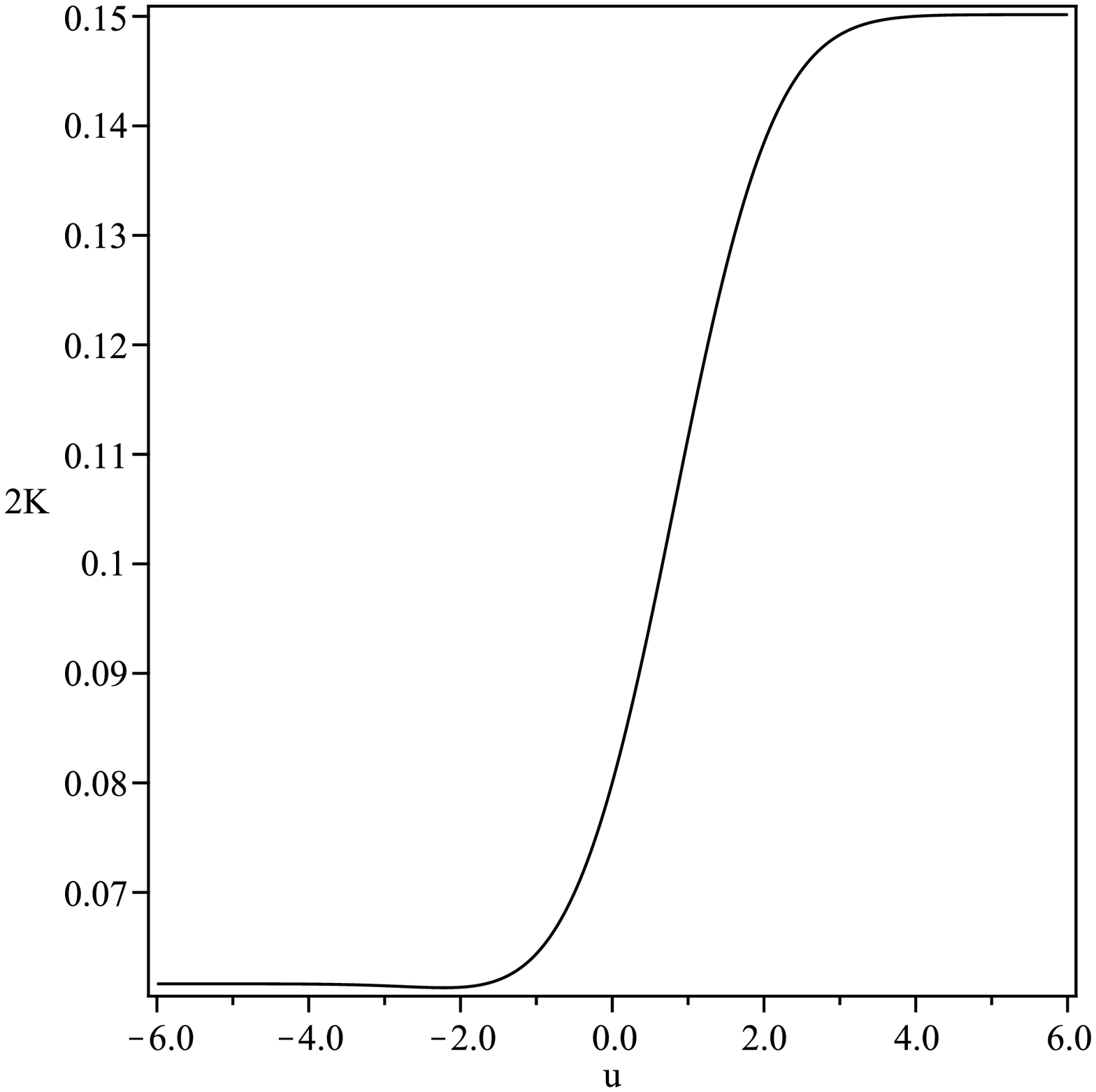}
\caption{Case 3- decreasing of the kinetic energy - initial conditions 
(\ref{10})}
\end{figure}

\section{Final Comments}

It is reasonable to consider that gravitational waves interact with the 
detector, which is inextricably linked with the space-time geometry.
The effect of this interaction results in some form of the memory 
effect as, for instance, in the velocity memory effect. The absorption of
energy by a system of free particles, after the passage of a gravitational wave,
was already envisaged by Bondi \cite{Bondi}. This is the physical feature one 
would expect from our experience with electromagnetic waves and charged 
classical particles.

The absorption of energy {\it from} a system of free particles, by a non-linear
gravitational wave, is an unexpected feature. In view of this result,
the variation of kinetic energy of free particles is not, in principle, a 
useful criterium for investigating the memory effect, since both kinds of
variation occur for the same wave. However, we have not 
been able to classify the conditions under which the kinetic energy of the 
particles increases or decreases. This issue will be investigated elsewhere.

\end{document}